# A vector magnetometer based on a single spin-orbit torque anomalous Hall device

Xin Chen, Hang Xie, Haoxuan Shen, and Yihong Wu[*]

*Department of Electrical and Computer Engineering, National University of Singapore, Singapore 117576*

In many applications, the ability to measure the vector information of a magnetic field with high spatial resolution and low cost is essential, but it is still a challenge for existing magnetometers composed of multiple sensors. Here, we report a single-device based vector magnetometer, which is enabled by spin-orbit torque, capable of measuring a vector magnetic field using the harmonic Hall resistances of a superparamagnetic ferromagnet (FM)/heavy metal (HM) bilayer. Under an ac driving current, the first and second harmonic Hall resistances of the FM/HM bilayer show a linear relationship with the vertical and longitudinal component (along the current direction) of the magnetic field, respectively. By employing a L-shaped Hall device with two orthogonal arms, we can measure all the three field components simultaneously, so as to detect both the amplitude and direction of magnetic field in a three-dimensional space. As proof of concepts, we demonstrate both angular position sensing on the three coordinate planes and vector mapping of magnetic field generated by a permanent magnet, both of which are in good agreement with the simulation results. Crosstalk between vertical and longitudinal field components at large field is discussed using theoretical models.

[*] E-mail: elewuyh@nus.edu.sg

# I. INTRODUCTION

Advancements in magnetic sensing have contributed immensely to a wide range of scientific and technological fields from fundamental physics, chemistry and biology to practical applications such as data storage and medical imaging, but measurement of a vector field with high spatial resolution using a single magnetic sensor remains to be a challenge. Compact and low-cost magnetometers such as Hall and magnetoresistance (MR) sensors are readily available [1-7], but these sensors only detect the magnetic field in a specific direction. To detect the field components in three orthogonal directions in space simultaneously, the common method is to integrate three magnetic sensors whose detection axes are perpendicular to each other [8-14] or to use magnetic flux guide to change the direction of one of the field components [12,15,16], as shown schematically in Fig. 1(a) and 1(b). However, these techniques often suffer from either, or a combination of, high cost, large size, low spatial resolution, high noise, and crosstalk among the measurement axes. The recently reported nitrogen-vacancy magnetometer does provide good spatial resolution, but it requires sophisticated optics and expensive microwave source to operate, making it not suitable for cost-sensitive applications [17-19].

In this work, we propose and experimentally demonstrate a high spatial resolution and low-cost vector magnetometer, which we term it as harmonic Hall vector magnetometer, based on a single Hall device enabled by spin-orbit torque (SOT) [20]. The sensor has an extremely simple structure, which consists of just an L-shaped (Co, Fe)B/Ta Hall bar with two mutually perpendicular arms (to facilitate discussion, hereafter we refer them to as arm-X and arm-Y), as shown in Fig. 1(c). The thickness of (Co, Fe)B is optimized to make it superparamagnetic such that it exhibits a perpendicular magnetic anisotropy (PMA) with negligible hysteresis, as confirmed by anomalous Hall effect (AHE) measurements. We apply an ac current to the Hall device and measure the first (1st) and second (2nd) harmonic components of the AHE signal from both arms. At small field, the former is proportional to the out-of-plane (OP) component ($H_z$) of the external field whereas the latter is proportional to the in-plane (IP) component ($H_x$ or $H_y$) along the driving current direction due to damping-like (DL) SOT effective field. As current directions are perpendicular to each other in the two arms of the L-shaped Hall device, one can simultaneously determine $H_x$ and $H_y$ from the 2nd harmonic AHE signal of the respective arms and $H_z$ from the 1st harmonic AHE signal of either arm. In this way, we realize precise measurement of the three field components simultaneously using a single device without any post-measurement data processing. The linearity ranges for the in-plane and out-of-plane components are ±100 Oe and ±50 Oe, respectively, which make the sensor suitable for a wide range of applications. As proof-of-concept applications, we demonstrate both angular position detection and vector field mapping using the developed sensor. The average angle error across 360° is less than 1° in the three Cartesian coordinate planes, and the vector mapping of magnetic field generated by a cylindrical magnet agrees well with the simulation results. Crosstalk between vertical and longitudinal field components at large field is discussed using theoretical models.

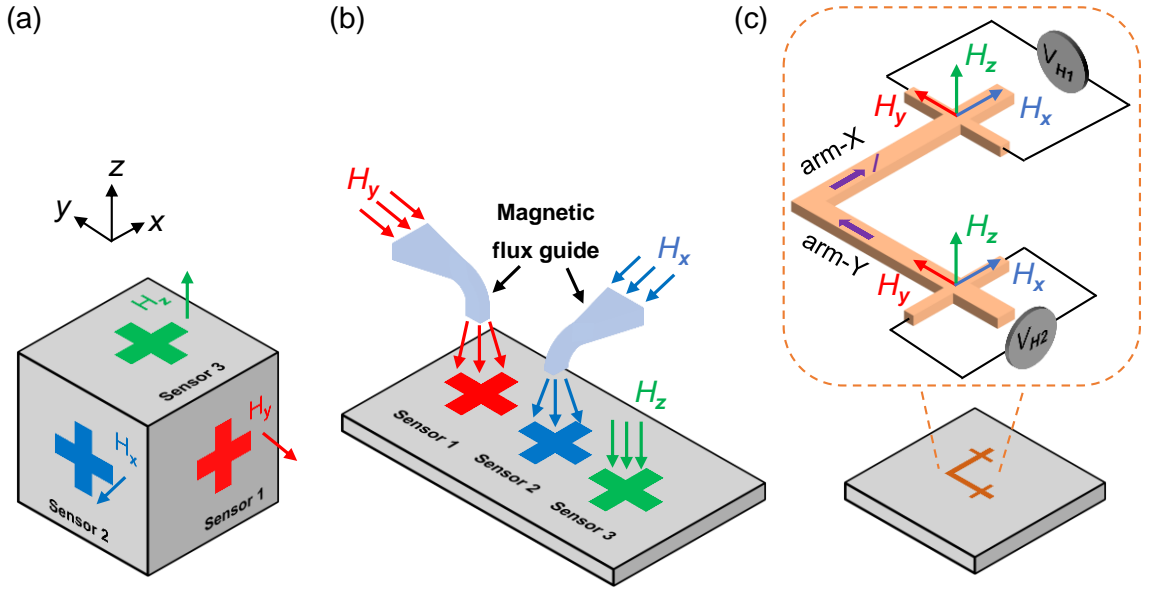

FIG. 1. Schematics of (a) vector magnetometer using multiple sensors; (b) vector magnetometer with multiple sensors and magnetic flux guide; (c) vector magnetometer presented in this work. Purple arrows indicate the current directions.

## II. OPERATION PRINCIPLE AND EXPERIMENTAL DETAILS

### A. Principe of harmonic Hall vector magnetometer

In general, the hysteresis (or M-H) loop of ferromagnetic materials may be modelled using the hyperbolic analytical approximation of the Everett integral based on the stochastic Preisach approach [21,22]. According to this model, the ascending $M_a$ and descending $M_d$ branches of the M-H loop may be expressed as

$$M_a = M_s \tanh\left[\frac{1}{H_0}(H - H_c)\right] + F(H_m), \tag{1}$$

$$M_d = M_s \tanh\left[\frac{1}{H_0}(H + H_c)\right] - F(H_m), \tag{2}$$

where $M_s$ is the saturation magnetization, $H_c$ is coercivity, $H_m$ is the maximum excitation field, $1/H_0$ is the differential permeability at $H = H_c$, and $F(H_m) = (M_S/2)\{\tanh[(H_m + H_c)/H_0] - \tanh[(H_m - H_c)/H_0]\}$. When the coercivity is negligible and the loop is symmetrical at large fields, i.e., $H_c = 0$ and $F(H_m) = 0$, $M_a$ and $M_d$ can be written as

$$M_a = M_d = M_s \tanh\frac{H}{H_0}. \tag{3}$$

Thin films with such kind of magnetic properties have been employed to realize superparamagnetic tunnel junctions [23-27]. In the present case, the L-shaped device consists of a superparamagnetic (Co, Fe)B/Ta bilayer with weak PMA and negligible hysteresis, and we use it to detect the vector information of magnetic field. When the bilayer is subjected to both an external magnetic field along z-axis (*i.e.*, easy axis) and an in-plane current along x-axis, the vertical component of the magnetization can be written as $M_z = M_S \tanh(H_z^{eff}/H_0)$, where $H_z^{eff}$ is the effective magnetic field along z-axis, including both external field ($H_z$) and the DL SOT effective field, i.e., $H_z^{eff} = H_z + H_z^{DL}$. The $H_z^{DL}$ is known to be proportional to the projection of magnetization along current direction, i.e., $H_z^{DL} = H^{DL} m_x$ with $H^{DL}$ the magnitude of DL effective field and $m_x$ the normalized magnetization in x-direction [20]. The $H_z^{DL}$ functions as an effective "knob" to detect longitudinal field $H_x$ as when $H_x$ is small, $H_z^{DL} \approx (H^{DL}/H_k^{eff})H_x$, where $H_k^{eff}$ is the effective anisotropy field [28-33]. Since the (Co, Fe)B/Ta bilayer exhibits PMA, the anomalous resistance $R_H$ can be written as

$$R_H = R_0 + R_{AHE}\tanh\left[\frac{1}{H_0}\left(H_z + \frac{H^{DL}}{H_k^{eff}}H_x\right)\right], \tag{4}$$

where $R_{AHE}$ is AHE resistance at saturation and $R_0$ is the offset resistance induced by misalignment of Hall voltage electrodes.

When the sensor is driven by an ac current $I = I_0 \sin \omega t$, the DL effective field can be written as $H_{DL} = (\hbar/2e)[\theta_{SH}/(M_S t_{FM} S)]I_0 \sin \omega t$, here, $\theta_{SH}$ is the effective spin Hall angle of Ta, $t_{FM}$ is the thickness of (Co, Fe)B layer, $S$ is the cross-section area of the device, $I_0$ and $\omega$ are the amplitude and angular frequency of the ac current, respectively, $\hbar$ is the reduced Planck constant, and $e$ is the electron charge [33]. From Eq. (4), we can obtain the Hall voltage for the Hall cross of arm-X as

$$V_H = I_0 \sin \omega t \, R_0 + I_0 \sin \omega t \, R_{AHE} \tanh\left[\frac{1}{H_0}\left(H_z + \frac{\hbar}{2e}\frac{\theta_{SH}}{M_S t_{FM} S}\frac{H_x}{H_k^{eff}}I_0 \sin \omega t\right)\right]$$

$$= I_0 \sin \omega t \, R_0 + I_0 \sin \omega t \, R_{AHE} \tanh\left[\frac{1}{H_0}(H_z + AH_x I_0 \sin \omega t)\right]$$

$$= I_0 R_0 \sin \omega t + I_0 R_{AHE} \sin \omega t \sum_{n=1}^{\infty} \frac{2^{2n}(2^{2n}-1)B_{2n}\left[\frac{1}{H_0}(H_z + AH_x I_0 \sin \omega t)\right]^{2n-1}}{(2n)!}, \quad (5)$$

where $A = (\hbar/2e)[\theta_{SH}/(M_s t_{FM} S H_k^{eff})]$ and $B_n$ is $n$th Bernoulli number. When both $H_z$ and $H_x$ are small, terms with $n \geq 2$ in Eq. (5) are negligible and Eq. (5) can be reduced to

$$V_H \approx \left(I_0 R_0 + \frac{I_0 R_{AHE}}{H_0}H_z\right)\sin \omega t + \left(I_0^2 \frac{R_{AHE}}{H_0}AH_x\right)(\sin \omega t)^2$$

$$= \frac{1}{2}I_0^2 \frac{R_{AHE}}{H_0}AH_x + I_0\left(R_0 + \frac{R_{AHE}}{H_0}H_z\right)\sin \omega t - \frac{1}{2}I_0^2 \frac{R_{AHE}}{H_0}AH_x \cos 2\omega t. \quad (6)$$

As can be seen from Eq. (6), the amplitudes of the 1st and 2nd harmonic $V_H$ are linearly proportional to $H_z$ and $H_x$, respectively, which facilitates the discrimination of $H_z$ and $H_x$ contributions to the output signal. The same results also apply to arm-Y by simply replacing $H_x$ with $H_y$. By doing so, we can detect $H_x$, $H_y$, and $H_z$ simultaneously using a single device.

### B. Sample preparation and experimental methods

Stack of MgO(1.1)/(Co, Fe)B($t_{CoFeB}$)/Ta(1.1)/MgO(2)/Ta(1.5) (the numbers in parentheses indicate the layer thickness in nm) thin films were deposited on the Si/SiO$_2$ substrates by magnetron sputtering with a base pressure of $1 \times 10^{-8}$ Torr and a working pressure of $3 \times 10^{-3}$ Torr. Here, $t_{CoFeB}$ is the thickness of the (Co, Fe)B layer. The Mircotech laserwriter system with a 405 nm laser was used to directly expose the photoresist (Mircoposit S1805), after which it was developed in MF319 to form the L-shaped Hall bar pattern. After film deposition, the photoresist was removed by a mixture of Remover PG and acetone to complete the Hall device fabrications. The processes of photography and lift-off were repeated to form electrodes and contact pads with the layers of Ta(5)/Cu(150)/Pt(10) for Hall bars. Finally, the devices were all annealed at 250°C for 1 hour with a pressure $< 1 \times 10^{-5}$ Torr. The electrical measurements were performed in the Quantum Design VersaLab PPMS with a sample rotator. The ac or dc current was applied by Keithley 6221 current source. The Hall voltage was measured by the Keithley 2182 nanovoltmeter (for dc voltage) and the 500 kHz MFLI lock-in amplifier from Zurich Instrument (for harmonic voltages). Vector mapping of magnetic field generated by a permanent magnet was performed using a xy stage in a normal experimental room.

## III. RESULTS AND DISCUSSION

### A. Thickness optimization and current-induced switching of MgO/(Co, Fe)B/Ta

Prior to the device fabrication, the thickness of (Co, Fe)B film has been optimized to reduce the coercivity to nearly zero [34-36]. Figure 2(a) shows the anomalous Hall effect (AHE) loops for MgO(1.1)/(Co, Fe)B($t_{CoFeB}$)/Ta(1.1)/MgO(2)/Ta(1.5) multilayers with different (Co, Fe)B thicknesses ($t_{CoFeB}$ = 1.2, 1.4 and 1.6) at room temperature. As can be seen, the sample with 1.4 nm (Co, Fe)B exhibits an AHE loop with sizable AHE resistance and negligible hysteresis, whereas the sample with 1.2 nm (Co, Fe)B exhibits a significantly decreased AHE and the sample with 1.6 nm (Co, Fe)B exhibits an AHE loop with a hysteresis, which is undesired for a linear field sensor. Therefore, 1.4 nm (Co, Fe)B exhibited the superparamagnetic behaviour at room temperature and it was used as a FM layer in the developed sensor. In typical AHE measurements with small current, $H^{DL}$ can be ignored and Eq. (4) can be written as $R_H = R_0 + R_{AHE}\tanh[H_z/H_0]$, which fits well the measured AHE curve in MgO(1.1)/(Co, Fe)B(1.4)/Ta(1.1)/MgO(2)/Ta(1.5) with the parameters $R_{AHE} = 16.24$ Ω, $R_0 = 0.09$ Ω, and $H_0 = 49.60$ Oe, as shown in Fig. 2(b). The results confirm the the validity of Eq. (4) for describing the AHE of the bilayer structure used in this study.

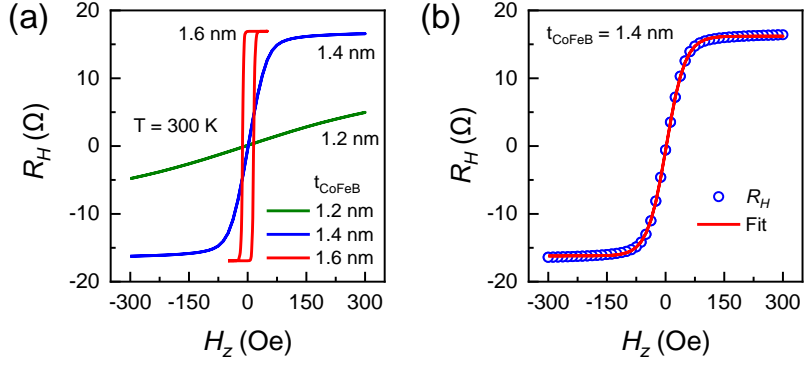

FIG. 2. (a) AHE loops for MgO(1.1)/(Co, Fe)B($t_{CoFeB}$)/Ta(1.1)/MgO(2)/Ta(1.5) multilayers with $t_{CoFeB}$ = 1.2 (green), 1.4 (blue) and 1.6 (red). (b) AHE loop of MgO(1.1)/(Co, Fe)B(1.4)/Ta(1.1)/MgO(2)/Ta(1.5) at a dc current 1mA and at room temperature. Measured data (circle) is fitted with Eq. (4) (solid-line).

After thickness optimization, we proceeded to fabricate the L-shaped device as shown schematically in Fig. 1(c). The width and length of each arm are 15 µm and 120 µm, respectively. Figure 3(a) shows the anomalous Hall resistance ($R_H$) in arm-X and arm-Y as a function of external field $H_z$ measured at an applied dc current of 1 mA for MgO(1.1)/(Co, Fe)B(1.4)/Ta(1.1)/MgO(2)/Ta(1.5). As can be seen, both arms of the device exhibit AHE with the coercivity of 0 Oe and $R_{AHE}$ of 16.2 Ω at room temperature, indicating good film uniformity in the whole device. Furthermore, Fig. 3(b) shows the current-induced switching loops of arm-X with different $H_x$ at room temperature. As can be seen, the switching loops corroborate well with the SOT mechanism. When $H_x = 0$, no current-induced switching occurs. The switching ratio increases with increasing $H_x$, and the switching polarity is determined by the directions of both $I_x$ and $H_x$. The switching saturates at 3.85 mA. The current-induced Hall resistance difference $\Delta R_H$ between ±3.85 mA at different $H_x$ is summarized in Fig. 3(c) (circle), which can be well fitted with $\Delta R_H = \Delta R_0 \tanh\{I_0(1/H_0)(\hbar/2e)[\theta_{SH}/(M_s t_{FM} SH_k^{eff})]H_x\}$ (solid-line), where $\Delta R_0 = 4.92$ Ω, $\theta_{SH} = -0.07$, $I_0 = 4.05$ mA, $M_s = 500.1$ emu/cm$^3$, and $H_k^{eff} = 699.91$ Oe. This agrees well with Eq. (4) with $H_z = 0$.

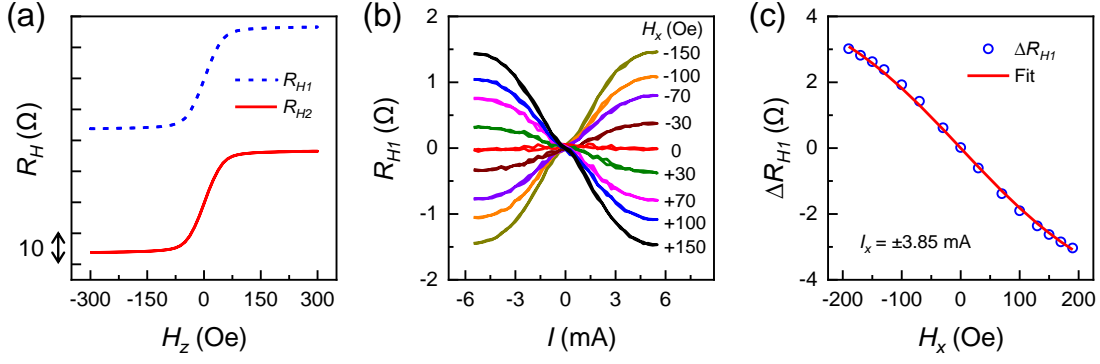

FIG. 3. (a) AHE loops of arm-X (upper curve) and arm-Y (lower curve) measured at a dc current of 1 mA at room temperature. (b) Current-induced switching loops of arm-X measured at different assistive fields, $H_x$. (c) Current-induced Hall resistance difference at ±3.85 mA with different $H_x$ (circle: experiment; solid-line: fitting).

### B. Measurements of individual field component

To demonstrate the proof-of-concept operation of the harmonic Hall vector magnetometer, we firstly conducted the harmonic Hall measurements on the L-shaped device when the external magnetic field was swept along three orthogonal axes separately. Harmonic Hall resistance is defined by the harmonic Hall voltage divided by the applied current amplitude. Figure 4 shows the 1st and 2nd harmonic Hall resistance as a function of sweeping fields in x-, y-, and z-direction, respectively. The device was driven by an ac current with an amplitude of 4 mA and frequency of 115 Hz. The 1st and 2nd harmonic Hall voltages were acquired using the lock-in amplifier. Shown in Fig. 4(a) are the harmonic Hall resistance, $R_{H1}^{\omega}$ and $R_{H1}^{2\omega}$, of arm-X when the field is swept along z- direction (note: zero-field offset has been subtracted out). As can be seen, $R_{H1}^{\omega}$ is linear to the $H_z$ at small field and saturates at high field, while the amplitude of $R_{H1}^{2\omega}$ is almost zero in the entire field range. Figure 4(d) displays $R_{H1}^{\omega} - H_z$ in a smaller range from -50 Oe to +50 Oe. Within this range, the curve shows good linearity with the maximum linearity error less than 3%, negligible hysteresis and a sensitivity of 149.44 mΩ/Oe. An opposite trend is obtained when the field sweeps in x-direction, as shown in Fig. 4(b). In this case, $R_{H1}^{2\omega}$ is linear with respect to $H_x$ at small field and saturates at high field. Although a small $R_{H1}^{\omega}$ is also observed, it could be due to misalignment of field in this measurement. As can be seen from Fig. 4(e), $R_{H1}^{2\omega}$ exhibits good linearity with maximum linearity error less than 3%, negligible hysteresis, and a sensitivity of 3.36 mΩ/Oe in the field range of -100 Oe to +100 Oe. The much smaller sensitivity compared to $R_{H1}^{\omega}$ obtained by sweeping the field in z-direction is expected as it is a 2nd order effect. Similar results were obtained for arm-Y, as shown in Fig. 4(c) and 4(f). In this case, the field was swept in y-direction. As expected, $R_{H2}^{2\omega}$ is linear to $H_y$ at small field and saturates at large field, while $R_{H2}^{\omega}$ is nearly zero in the entire field range. The sensitivity, 3.30 mΩ/Oe, and linear range, -100 Oe to +100 Oe, are similar to those of arm-X, indicating good uniformity in both the film stack and patterned device. The above results show that the L-shaped device functions well as a linear sensor when there is only one field component present.

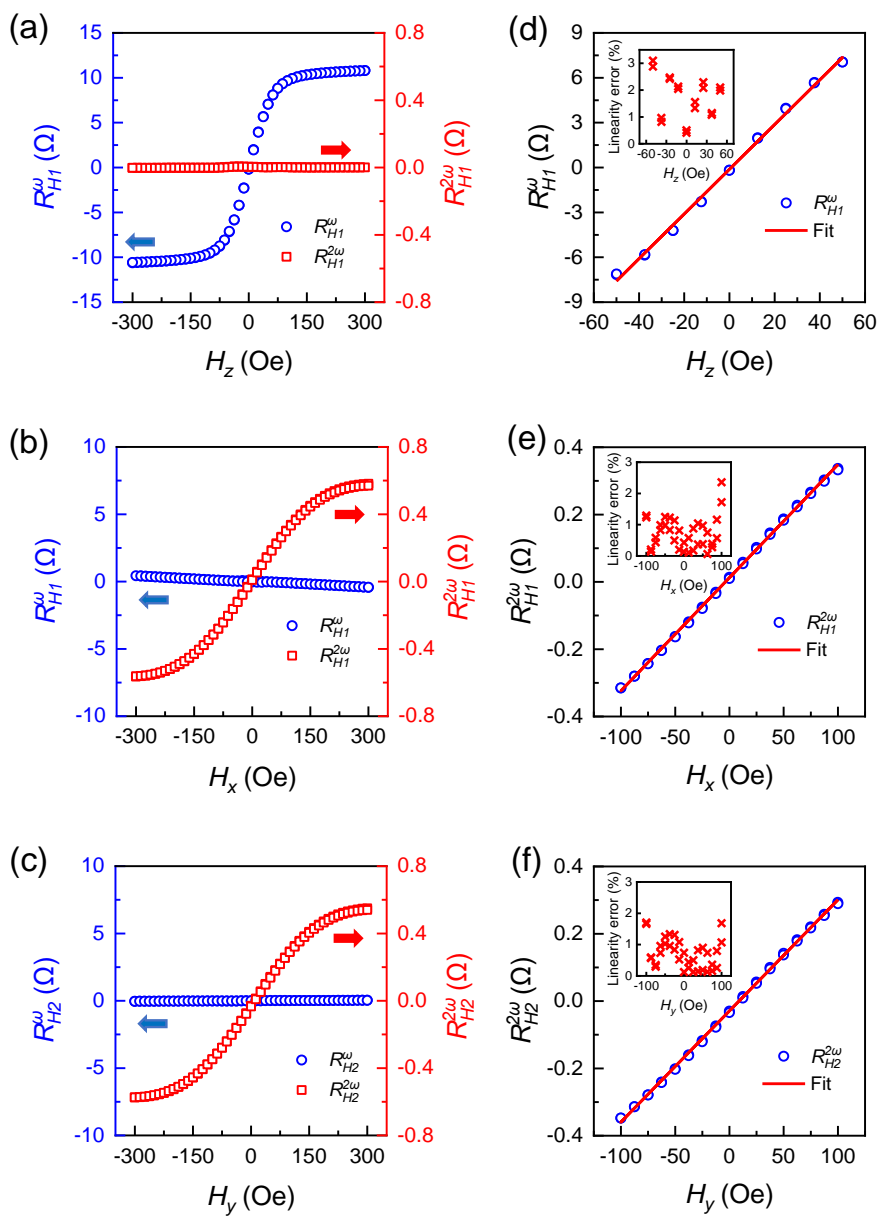

FIG. 4. (a)-(c) 1st (circle) and 2nd (square) harmonic Hall resistance with the field swept in (a) z-direction (arm-X), (b) x-direction (arm-X) and (c) y-direction (arm-Y), respectively. (d)-(f) Harmonic Hall resistance as a function of external field $H_x$, $H_y$, and $H_z$, respectively, in the small field range. Solid-lines are linear fittings with the linearity error given in the insets.

### C. Angle detection on three coordinate planes

Next, we examine the possibility of using the L-shaped device as a biaxial sensor. Due to unavailability of vector electromagnet, here we use the Hall device to determine the direction of a magnetic field with constant magnitude but with its direction rotating in the three coordinate planes. As shown schematically in Figs. 5(a)-(c), when the field rotates in zx, yz and xy planes, the 1st and 2nd harmonic Hall resistance are given by i) rotation in zx plane: $R_{H1}^{\omega} = R_0 + R_{AHE} H \cos\theta_{zx}^H / H_0$, $R_{H1}^{2\omega} = -(I_0 R_{AHE} A H \sin\theta_{zx}^H)/(2H_0)$ ; ii) rotation in yz plane: $R_{H2}^{\omega} = R_0 + R_{AHE} H \sin\theta_{yz}^H / H_0$, $R_{H2}^{2\omega} = -(I_0 R_{AHE} A H \cos\theta_{yz}^H)/(2H_0)$; and iii) rotation in xy plane: $R_{H1}^{2\omega} = -(I_0 R_{AHE} A H \cos\theta_{xy}^H)/(2H_0)$, $R_{H2}^{2\omega} = -(I_0 R_{AHE} A H \sin\theta_{xy}^H)/(2H_0)$. Here, $H$ is the external magnetic field amplitude, $\theta_{zx}^H$, $\theta_{yz}^H$ and $\theta_{xy}^H$ are the angles between the rotating field and z, y, x axes, respectively, on the zx, yz and xy plane. The angle is positive when the rotation direction and axis follows the right-handed rule.

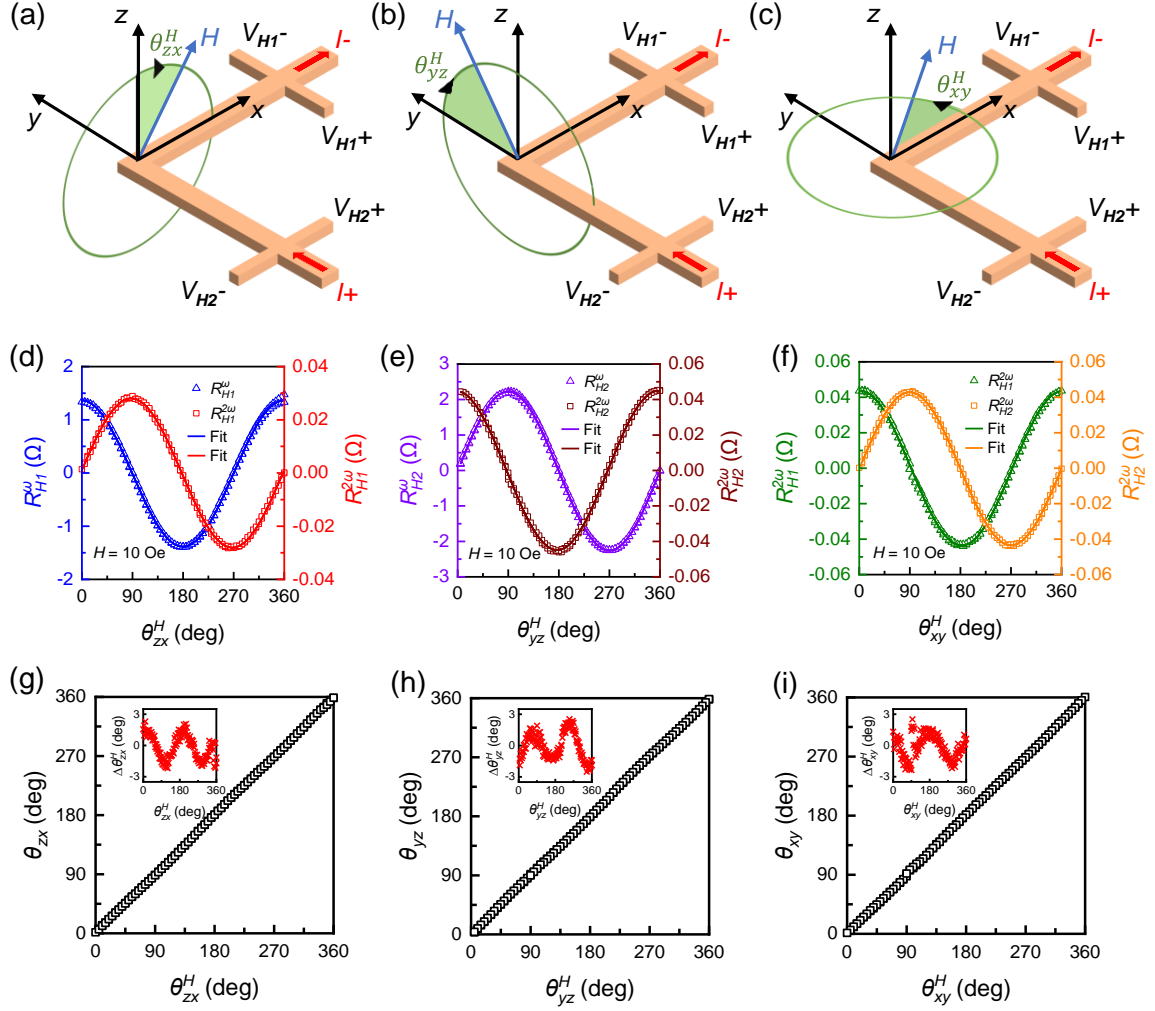

FIG. 5. (a)-(c) Measurement geometries with the field rotating in zx, yz and xy planes, respectively. (d) and (e) 1st and 2nd harmonic Hall resistance of arm-X with the field rotating in the zx and yz planes, respectively. (f) 2nd harmonic Hall resistance of arm-X and arm-Y with the field rotating in the xy plane. The field strength is fixed at 10 Oe. (g)-(i) Actual field angle ($\theta_{ij}^H$) versus calculated field angle ($\theta_{ij}$) in the full range of 360° when the field rotates in zx, yz and xy planes, respectively (i, j = x, y, z). The insets show the angle errors.

Figure 5(d) shows $R_{H1}^\omega$ and $R_{H1}^{2\omega}$ as a function of $\theta_{zx}^H$ from 0° to 360° when the device is driven by an ac current with an amplitude of 4 mA and frequency of 115 Hz. The external field strength is 10 Oe. Both signals were acquired from the single device simultaneously using the lock-in amplifier. As expected, $R_{H1}^\omega - \theta_{zx}^H$ curve is in a cosine shape while $R_{H1}^{2\omega} - \theta_{zx}^H$ follows a sine function. They can be fitted well with $R^{\omega}_{H10}\cos\theta^H_{zx}$ (solid-line in blue) and $R^{2\omega}_{H10}\sin\theta^H_{zx}$ (solid-line in red), respectively, where $R^{\omega}_{H10}$ and $R^{2\omega}_{H10}$ are the amplitudes of $R^{\omega}_{H1}-\theta^H_{zx}$ and $R^{2\omega}_{H1}-\theta^H_{zx}$ curves, respectively. Similar results are obtained for the field rotating in yz and xy planes, as shown in Fig. 3(e) and 3(f), respectively. The field strength remains to be 10 Oe. With the sine and cosine dependence of the harmonic Hall resistance, we can calculate the field angle as $\theta_{zx} = \mathrm{atan2}(-R^{\omega}_{H1}/R^{\omega}_{H10}, -R^{2\omega}_{H1}/R^{2\omega}_{H10}) + \pi$, $\theta_{yz} = \mathrm{atan2}(-R^{2\omega}_{H2}/R^{2\omega}_{H20}, -R^{\omega}_{H2}/R^{\omega}_{H20}) + \pi$, and $\theta_{xy} = \mathrm{atan2}(-R^{2\omega}_{H1}/R^{2\omega}_{H10}, -R^{2\omega}_{H2}/R^{2\omega}_{H20}) + \pi$. Figures 5(g)-(i) show the relationship between the detected angle and actual field angle on the three coordinate planes. As can be seen, the detected angle is almost the same as the actual angle. As shown in the insets, the maximum angle error is around 3°, and the average error from 0 to 360° is less than 1°.

Next, we turn to the field angle detections at large magnetic fields. Figure 6(a) shows $R^{\omega}_{H1}$ and $R^{2\omega}_{H1}$ as a function of $\theta^H_{zx}$ from 0° to 360° when the external field strength is 30 Oe, whereas Fig. 6(b) shows $R^{2\omega}_{H1}$ and $R^{2\omega}_{H2}$ as a function of $\theta^H_{xy}$ from 0° to 360° when the external field strength is 50 Oe. The corresponding relationships between the detected angle and actual field angle are shown in Fig. 6(c) and 6(d), respectively with the angle error given in the insets. As can be seen, the maximum angle error increases to be 8° for $\theta^H_{zx}$ with an external field strength of 30 Oe. As can be seen from Fig. 6(a), the measured $R^{\omega}_{H1}$ and $R^{2\omega}_{H1}$ curves deviate from the cosine and sine fitting curves, which results in a larger angle error. But the angle error for $\theta^H_{xy}$ remains to be less than 3° with an external field strength of 50 Oe, as shown in Fig. 6(b) and 6(d). The main reason for the larger angle error, especially for $\theta^H_{zx}$ (and $\theta^H_{yz}$), is the crosstalk between vertical and longitudinal field components induced by the non-negligible higher order effect at large fields (see Appendix for details). The error can be reduced by removing the higher order effect, as shown in Fig. 7.

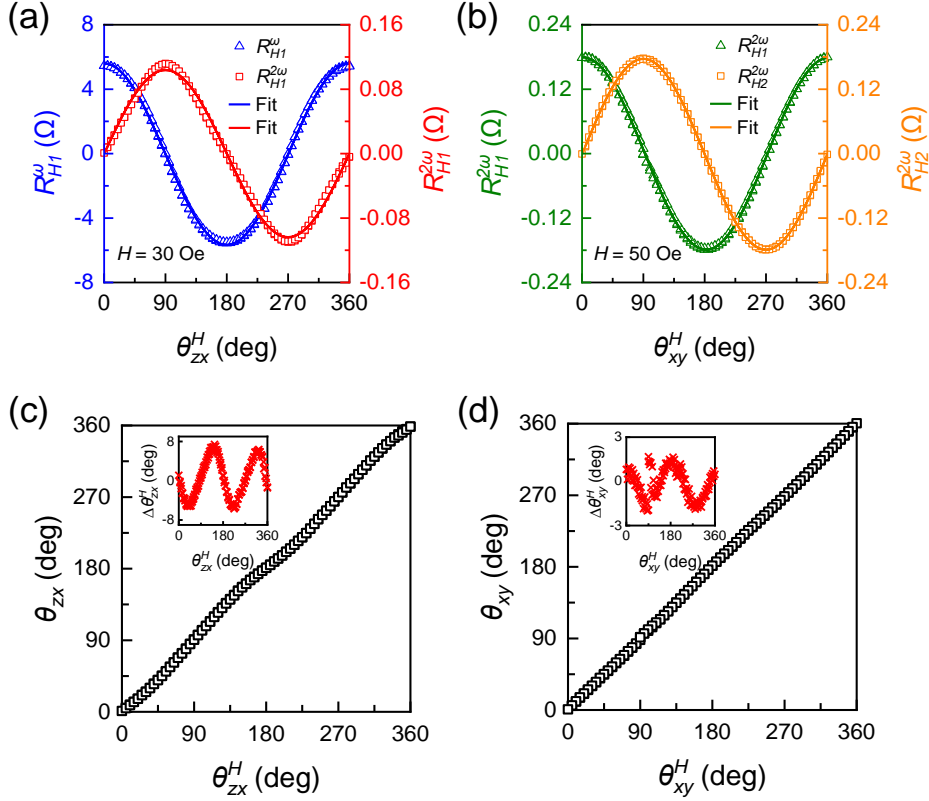

FIG. 6. (a) 1st and 2nd harmonic Hall resistance of arm-X with a rotating field of 30 Oe in the zx plane. (b) 2nd harmonic Hall resistance of arm-X and arm-Y with a rotating field of 50 Oe in the xy plane. The plots of actual field angle ($\theta_{ij}^H$) versus calculated field angle ($\theta_{ij}$) in the full range of 360° (i, j = x, y, z) corresponding to (a) and (b) are shown in (c) and (d), respectively. The insets are the angle errors.

Figure 7(a) shows the $R_{H1}^\omega$ and $R_{H1}^{2\omega}$ as a function of $\theta_{zx}^H$ from 0° to 360° when the external field strength is 30 Oe and fitting results using $R_H^\omega = R_0 + k_1 \cos\theta_{zx}^H + k_2 \cos 3\theta_{zx}^H$ and $R_H^{2\omega} = k_3 \sin\theta_{zx}^H + k_4 \sin 3\theta_{zx}^H$. Here, $k_1 = \frac{R_{AHE}}{H_0}H - \frac{R_{AHE}}{4H_0^3}H^3 - \frac{3I_0^2 R_{AHE} A^2}{16H_0^3}H^3$, $k_2 = \frac{3I_0^2 R_{AHE} A^2}{16H_0^3}H^3 - \frac{R_{AHE}}{12H_0^3}H^3$, $k_3 = -\frac{I_0 R_{AHE} A}{2H_0}H + \frac{I_0^3 R_{AHE} A^3}{8H_0^3}H^3 + \frac{I_0 R_{AHE} A}{8H_0^3}H^3$ and $k_4 = \frac{I_0 R_{AHE} A}{8H_0^3}H^3 - \frac{I_0^3 R_{AHE} A^3}{24H_0^3}H^3$ (See Appendix for details). As can be seen, $R_{H1}^\omega$ and $R_{H1}^{2\omega}$ can be well fitted with $R_0 = 0.038\ \Omega$, $k_1 = 5.632\ \Omega$, $k_2 = -0.154\ \Omega$, $k_3 = 0.104\ \Omega$ and $k_4 = -0.006\ \Omega$. $k_2 \cos 3\theta_{zx}^H$ in $R_H^\omega$ and $k_4 \sin 3\theta_{zx}^H$ in $R_H^{2\omega}$ are induced by the crosstalk between vertical and longitudinal field components. Next, we subtracted the $k_2 \cos 3\theta_{zx}^H$ and $k_4 \sin 3\theta_{zx}^H$ from $R_{H1}^\omega$ and $R_{H1}^{2\omega}$, respectively. As shown in Fig. 7(b), after the subtractions, $R_{H1}^\omega$ and $R_{H1}^{2\omega}$ can be well fitted with cosine and sine functions, respectively. Figure 7(c) shows the detected angle $\theta_{zx}$ which is calculated with $R_{H1}^\omega - k_2 \cos 3\theta_{zx}^H$ and $R_{H1}^{2\omega} - k_4 \sin 3\theta_{zx}^H$. As can be seen, the maximum angle error is less than 1° except for the few points near $\theta_{zx}^H = 360°$, which is presumably caused by the accuracy of sample rotator.

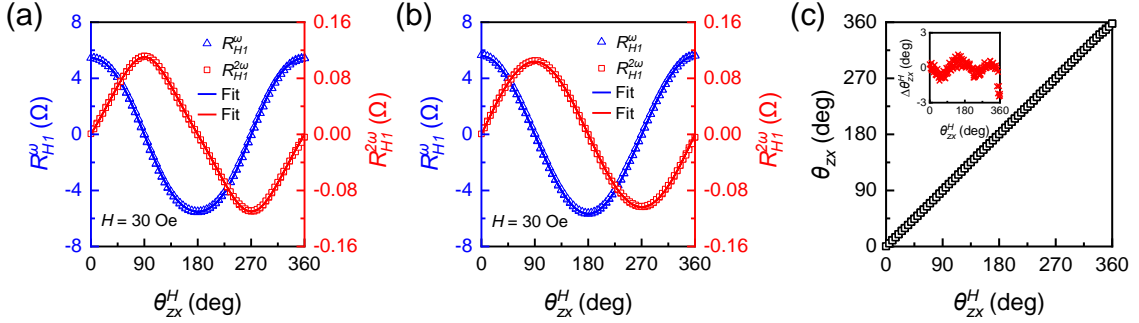

FIG. 7. (a) 1st and 2nd harmonic Hall resistance of arm-X with the 30 Oe field rotating in the zx plane. 1st and 2nd harmonic Hall resistance are fitted with Eq. (A4) (blue solid-line) and Eq. (A4) (red solid-line), respectively. (b) 1st and 2nd harmonic Hall resistance without $k_2 \cos 3\theta_{zx}^H$ and $k_4 \sin 3\theta_{zx}^H$ components, respectively. (c) Actual field angle ($\theta_{zx}^H$) versus calculated field angle ($\theta_{zx}$) in the full range of 360° when the field rotates in zx plane. The inset shows the angle error.

### D. Vector mapping of magnetic field generated by a permanent magnet

The above results show that the L-shaped device can function as both a single- and bi-axial sensor. To further demonstrate its capability as a vector magnetometer, we used the same device to map the field generated by a permanent magnet. Figure 8(a) shows the experimental setup where a cylindrical N35 permanent magnet ($B_s = 1.27$ T) with a diameter of 10 mm and thickness of 5 mm is attached to a non-magnetic fixture with its N-pole pointing down. The L-shaped Hall device was placed on an xy-stage right below the magnet with a distance of 33 mm from the bottom surface of the magnet and its center was aligned with that of the magnet. The X- and Y-arms of the device are aligned parallel with the two rails of the xy stage, and are indicated as x- and y-axis, respectively, in Fig. 8(a). As shown in Fig. 8(b), by scanning the sensor over an area of 50 mm × 12 mm, we successfully obtain the vector field distribution on a plane that is located at 33 mm below the magnet. The vectors are directly plotted from the field components, $H_x$, $H_y$, $H_z$, which were measured simultaneously using the Hall device through the harmonic Hall resistance $R_{H1}^{2\omega}$, $R_{H2}^{2\omega}$ and $R_{H1}^{\omega}$. To check the accuracy of the mapping results, we calculated the amplitude ($H$), and polar ($\theta_H$) and azimuthal ($\varphi_H$) angle of the field extracted from the measured field components, i.e., $H = (H_x^2 + H_y^2 + H_z^2)^{1/2}$, $\theta_H = \cos^{-1}(H_z/H)$, and $\varphi_H = \text{atan2}(-H_x/(H_x^2 + H_y^2)^{1/2}, -H_y/(H_x^2 + H_y^2)^{1/2}) + \pi$, and compared them with the simulation results. Figures 8(c)-(e) show the experimental data and the corresponding results simulated by the COMSOL Multiphysics® software are shown in Figs. 8(f)-(h), respectively. As can be seen, the measured field magnitude and angle are in good agreement with the simulation results. The results shown in Fig. 5 and Fig. 8 demonstrate clearly that the Hall device functions a vector magnetometer. It is worth pointing out that, due to the use of multiple sensors in commercial Hall vector magnetometers, the spatial distance between any two sensors is typically larger than 150 μm [37-40]. In contrast, in the device presented in this work, the distance between the two Hall cross is 70 μm and it can be further reduced to less than 30 μm or smaller (not shown here). The significant enhancement of spatial resolution will help to extend the application fields of vector magnetometer.

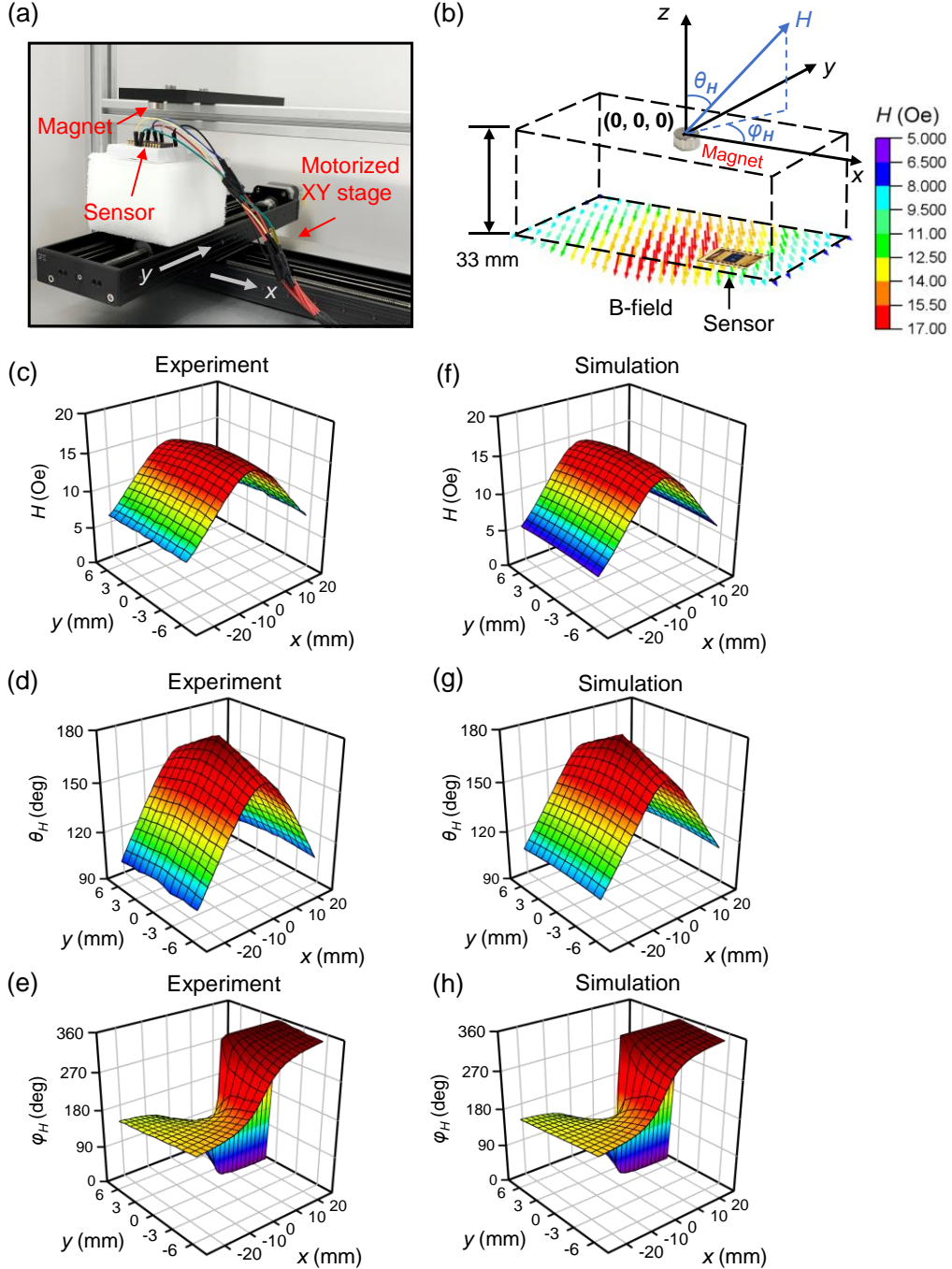

FIG. 8. (a) Experimental setup for vector mapping of magnetic field generated by a permanent magnet. (b) Measurement configuration according to the setup in (a). Also shown is the measured vector field distribution over an area of 50 mm × 12 mm on the xy plane. (c)-(e) Measured amplitude, polar and azimuthal angle of the magnetic field, respectively. (f)-(h) Simulated amplitude, polar and azimuthal angle of the magnetic field, respectively.

TABLE I. Comparison of SOT-based magnetic field sensors.

| Detection Mode | Detection range | Detection principle | Linear range | Remarks |
|---|---|---|---|---|
| Mode 1: Field along single axis | 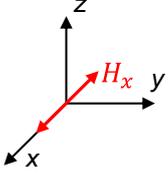 | SOT biasing and SMR | ±1 Oe | SMR sensor [41,42] |
| | | SOT-driven switching | ±10 Oe | STG sensor [33] |
| Mode 2: Field along a circle on xy plane (angle sensor) | 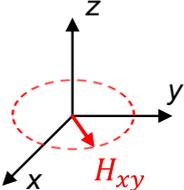 | Angular position sensing has been demonstrated using SMR sensor [41,42], STG sensor [33] and SOT driven Hall sensor [43] under a field strength of 1 Oe, 20 Oe, and 500 – 2000 Oe, respectively. | | |
| Mode 3: Field along 3 coordinate axes | 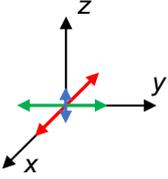 | Pulsed current-induced DW motion and detection | $H_x$: ±10 Oe<br>$H_y$: ±10 Oe<br>$H_z$: ±4 Oe | Measure $H_x$, $H_y$ and $H_z$, separately [44] |
| Mode 4: Field on three coordinate planes | 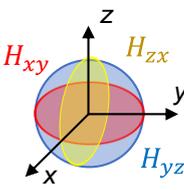 | SOT effective field and harmonic AHE | $H_x$: ±100 Oe<br>$H_y$: ±100 Oe<br>$H_z$: ±50 Oe<br>$H_{zx}$: 30 Oe<br>$H_{yz}$: 30 Oe<br>$H_{xy}$: 50 Oe | Pseudo-3D field mapping (this work) |
| Mode 5: Field in three-dimensional space | 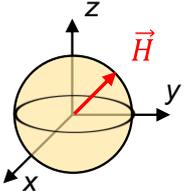 | SOT effective field and harmonic AHE | 20 Oe in all directions | Full 3D vector field mapping (this work) |

Before we conclude, we shall mention that several SOT-based magnetic sensors have been reported previously from both ours and other groups [33,41-44]. In Table I, we compare the detection mode, detection principle and field range of these sensors. The first type of sensor is a linear sensor which can only detect field along a single axis (Mode 1 in Table I). We have previously reported two types of SOT-enabled linear sensors, namely the spin Hall magnetoresistance (SMR) sensor for weak field detection [41,42], and the spin-torque gate (STG) sensor for intermediate range field detection [33]. Both types of sensors exhibit a field angle dependence similar to that of giant magnetoresistance sensor, and therefore they can be used for angle sensing as well (Mode 2 in Table I). In addition, we have also developed a SOT-based angular position sensor using two Hall crosses operating under pulsed current [43]. The sensor is able to detect the direction of a rotational field on the xy-plane, with a field strength from 500 Oe to 2000 Oe. Recently, R. Li et all. reported a SOT-based Hall device which can measure in-plane and out-of-plane fields separately (Mode 3 in Table I) in a relatively small field range [44]. The in-plane field detection is based on current-induced domain wall motion, and Joule heating is used to suppress the hysteresis in z-direction. It is important to note that separate measurement of individual field components is different from simultaneous measurements of three components because the latter has to deal with crosstalk among different field components. Crosstalk is unavoidable in SOT-based sensors because the SOT effective field is required for detecting the in-plane field components. The crosstalk causes the decrease of detectable field range for field along arbitrary directions in a three-dimensional space, which is absent when the field is only along a single coordinate axis (Mode 1 and 3 in Table I). As discussed in the appendix, the output signal of SOT-based anomalous Hall sensor may be written as $V \approx aH_z + bH_x + cH_x^2 H_z + dH_z^2 H_x + eH_z^3 + fH_x^3$ (when current is in x-direction) and $V \approx aH_z + bH_y + cH_y^2 H_z + dH_z^2 H_y + eH_z^3 + fH_y^3$ (when current is in y-direction). Here, a, b, c, d, e, f are field-independent constants. The 3rd and 4th terms are crosstalk terms which only appear when multiple field components are present. This aspect is overlooked in previous work as the measurement of a specific field component was performed without the presence of other field components. In contrast, here we have demonstrated a device that can measure all the three field components simultaneously, meaning that it can function in all detection modes listed in Table I, ranging from single axial to biaxial and triaxial sensing. In that sense, the harmonic Hall vector magnetometer represents the first fully functional vector magnetometer based on a single planar device.

## IV. CONCLUSIONS

In summary, we have proposed and demonstrated a fully functional single-device vector magnetometer enabled by the SOT and harmonic technique. The harmonic Hall vector magnetometer is an L-shaped

Hall device with two orthogonal arms. By measuring the 1st and 2nd harmonic Hall resistance of both arms, we can determine the three components of a vector field simultaneously. In addition to angle sensing on each coordinate plane, we have also shown that the proposed device is able to sense a vector field in any direction in three-dimensional space. Its simple configuration and high accuracy show its great potential in various fields requiring vector field measurements such as navigation, internet of things, smart electronics, and many other traditional and emerging applications.

**ACKNOWLEDGEMENTS**

This project is supported by the Ministry of Education, Singapore under its Tier 2 Grants (Grants No. MOE2018-T2-1-076 and No. MOE2017-T2-2-011) and ARTIC Grant (HFM-RP4) from National University of Singapore.

**APPENDIX: Crosstalk between vertical and longitudinal field components**

The results in Fig. 6 show that the angle error increases significantly at large field, especially for the angle detection on zx or yz plane. This is caused by the crosstalk induced by the non-negligible higher order terms at large field. When $H_z$ or $H_x$ is large, the term with $n = 2$ in Eq. (5) is non-negligible and it should be duly considered, i.e.,

$$V_H \approx I_0 R_0 \sin \omega t$$
$$+ I_0 R_{AHE} \sin \omega t \left\{ \frac{1}{H_0}(H_z + AH_x I_0 \sin \omega t) - \frac{1}{3}\left[\frac{1}{H_0}(H_z + AH_x I_0 \sin \omega t)\right]^3 \right\}. \quad (A1)$$

By expanding Eq. (A1), we can obtain the 1st and 2nd harmonic Hall resistance as follows:

$$R_H^\omega = R_0 + \frac{R_{AHE}}{H_0} H_z - \frac{1}{3}\frac{R_{AHE}}{H_0^3} H_z^3 - \frac{3}{4} I_0^2 R_{AHE} \frac{A^2}{H_0^3} H_z H_x^2, \quad (A2)$$

$$R_H^{2\omega} = -\frac{1}{2} I_0 \frac{R_{AHE}}{H_0} AH_x + \frac{1}{6} I_0^3 \frac{R_{AHE}}{H_0^3} A^3 H_x^3 + \frac{1}{2} I_0 \frac{R_{AHE}}{H_0^3} AH_z^2 H_x. \quad (A3)$$

In the case of a rotating field in xz plane, $H_x$ and $H_z$ can be written as $H_x = H \sin \theta_{zx}^H$ and $H_z = H \cos \theta_{zx}^H$, respectively, where $H$ is the field strength and $\theta_{zx}^H$ is the angle between the rotating field and z axis on the zx plane. Using the trigonometric identity, $R_H^\omega$ and $R_H^{2\omega}$ can be further written as

$$R_H^\omega = R_0 + k_1 \cos \theta_{zx}^H + k_2 \cos 3\theta_{zx}^H, \quad (A4)$$

$$R_H^{2\omega} = k_3 \sin \theta_{zx}^H + k_4 \sin 3\theta_{zx}^H, \quad (A5)$$

where $k_1 = \frac{R_{AHE}}{H_0} H - \frac{R_{AHE}}{4H_0^3} H^3 - \frac{3I_0^2 R_{AHE} A^2}{16 H_0^3} H^3$, $k_2 = \frac{3I_0^2 R_{AHE} A^2}{16 H_0^3} H^3 - \frac{R_{AHE}}{12 H_0^3} H^3$, $k_3 = -\frac{I_0 R_{AHE} A}{2H_0} H + \frac{I_0^3 R_{AHE} A^3}{8 H_0^3} H^3 + \frac{I_0 R_{AHE} A}{8 H_0^3} H^3$ and $k_4 = \frac{I_0 R_{AHE} A}{8 H_0^3} H^3 - \frac{I_0^3 R_{AHE} A^3}{24 H_0^3} H^3$. $A$ is defined in the main text. As can be seen from Eq. (A4) and Eq. (A5), the $k_2 \cos 3\theta_{zx}^H$ and $k_4 \sin 3\theta_{zx}^H$ induced by the higher order effect become non-negligible with a large magnetic field strength, resulting in the deviation of $R_H^\omega$ and $R_H^{2\omega}$ curve from the cosine and sine shape, respectively.